\documentclass[twocolumn,showpacs,superscriptaddress,prd]{revtex4}
\usepackage{amsmath}
\usepackage{amsfonts}
\usepackage{amssymb}
\usepackage{graphicx}
\usepackage{hyperref}
\usepackage{units}
\usepackage{color}%

\def\b{\begin{equation}}
\def\e{\end{equation}}
\def\ba{\begin{equation}\begin{array}{rl}}
\def\ea{\end{array}\end{equation}}

\def\la{\langle}
\def\ra{\rangle}
\def\be{\begin{align}}
\def\ee{\end{align}}

\begin{document}

\title{Effects of gravity on quantum key distribution}

\author{Roberto Pierini}
\email{robpierin@gmail.com}     
\affiliation{Institute of Theoretical Physics, University of Warsaw, 05-093 Warsaw, Poland}

\date{\today}

\begin{abstract}
The domain of quantum technologies has been recently broaden to satellites orbiting the earth. Long distances communication protocols cannot ignore the presence of the gravitational field and its effects on the quantum states. Here, we provide a general method to investigate how gravity affects the performance of various quantum key distribution protocols for continuous variable states of localized wave packets. We show that the consequences of gravity strongly depend on the configuration and the cryptographic scheme used.  
\end{abstract}



\maketitle

\section{Introduction}

Quantum key distribution (QKD) is one of the most important applications of quantum information theory. 
The two fundamental protocols, one by Bennet and Brassard \cite{BB84}, the other by Ekert \cite{E91}, have been extended to the realm of continuos variable (CV) systems and Gaussian states. The first protocol is based on a scheme where a state is prepared, sent and measured. The second is based on entanglement shared between the parties. It has been shown that the two schemes are equivalent \cite{BBM}. There are four standard protocols in Gaussian QKD: Alice prepares squeezed states with a Gaussian modulation along one of the two quadratures, sends them to Bob who measures them by performing homodyne detection \cite{CLV}; Alice prepares coherent states, Gaussian modulated along both quadratures, with Bob performing again homodyne detection \cite{GG}; a third is where Alice sends squezeed states and Bob applies now heterodyne detection \cite{Gar}; finally, the fourth protocol where Alice prepares coherent states and Bob heterodynes them \cite{Wee1}. 
Here, we refer mostly to the reasoning described in \cite{Gross}, where the authors address the problem of secrecy against individual attacks in the entanglement based (EB) scenario, and in his extension to collective attacks detailed in \cite{Gar}. Detailed reviews about QKD and Gaussian quantum information can be found in \cite{SBCDLP} and \cite{Wee}.

Performing such quantum protocols with satellites can allow for an application of such techniques to global scales. To account for all the physical phenomena that affect the photons is a difficult task and the effect of gravity must have some importance over long distance quantum communication, which is a working technology already \cite{sat1,sat2}. The effect of gravity can be taken into account as a curved dynamical background which modifies the equations of quantum field theory. This, has non-trivial effects even on fundamental phenomena such as the decay rate of unstable particles \cite{cloks}. Furthermore, gravity is locally equivalent to non-inertial motion according to the equivalence principle. Therefore, the effect of gravity on quantum states can be analogously studied as the effect of the acceleration of the observer on the quantum states.
While this is a well known fact since several decades \cite{CHM}, there have been some controversy during the years in the study of the Unruh effect in connection to standard quantum information tasks \cite{Dragan}. Infact, it has been clarified only recently how Gaussian states are properly affected by acceleration. The efforts have culminated in the work \cite{Ahm}, whose method has been succesfully extended \cite{RLDO} and applied to CV quantum information protocols \cite{GRKD}. Several previous works \cite{DRW,RW,ZSG} have meaningfully investigated the effect of the acceleration on a specific cryptographic scheme, where only one of the parties was allowed to accelerate, and the effect of gravity on information transmission \cite{Brus1,Brus2,MPW}. Here, we aim to extend those studies to various QKD schemes and to the general scenario in which both Alice and Bob are affected by acceleration or gravity. 

In Sec. \ref{protocol} we give a description of a general protocol for QKD with CV and his security condition in terms of key rates. 
In Sec. \ref{Unruh channel} is described how the covariance matrix of two localized wave packets changes when they are observed from a non inertial frame. In Sec. \ref{results} we study the influence of acceleration on the previously described protocol, in particular on the efficiency in distributing the secret key. 
Finally, we conclude with some comments in Sec. \ref{conclusion}.


\section{QKD with Continuous Variables}\label{protocol}

A standard protocol of CV QKD in the EB representation can be briefly described as follows. In the first step, Alice prepares a two mode squeezed vacuum state and sends one of the mode to Bob. Both Alice and Bob measure randomly either the $x$ or the $p$ quadrature of their share of the entangled state. After this, they perform the so-called sifting procedure where they drop the data which are outcomes of the measurements in different bases. 
Eve interacts with the quantum system to learn about the key. This action modifies the state in a way that can be observed. Therefore, in the next step the reliable parties reveal each other a randomly chosen sample of their data, in order to estimate the parameters of the channel and, consequently, the amount of information leaked to Eve. After sifting and parameters estimation, Alice and Bob share a string of correlated elements, called raw key.  
In the subsequent step they extract a common binary key from the data using classical communication. Depending on whether Alice's or Bob's data are used, the protocol is called direct reconciliation (DR) or reverse reconciliation (RR) respectively. Finally, they perform error correction and privacy amplification to get a perfectly correlated key. 

The standard procedure is to assume that Eve does a passive attack, where she replaces the real connection between Alice and Bob by a unitary operation that mimics the channel between them when we trace out Eve's modes. This operation is represented by a thermal noise channel of transmittance $T$ and noise referred to the input $\chi$. In practice, it can be implemented by a beam splitter of transmittance $T$, where the signal sent to Bob is combined with a thermal state of zero mean and variance $\la\hat X_{th}^2\ra=\frac{T}{1-T}\,\chi$, 
\begin{equation}\label{Eve}
\hat X_B=\sqrt{T}\,\hat X_A+\sqrt{1-T}\,\hat X_{th}  \,,
\end{equation}
where $\chi$ is the noise referred to the input of the channel
\begin{equation}
\chi=\frac{1-T}{T}+\varepsilon \,,
\end{equation}
and $\varepsilon$ the so-called excess noise. This attack is known as intercept and resend or cloning attack. 
To establish whether the set of binary symbols created in the protocols can be used in the encription task, the honest parties have to evaluate the so-called secret key rate $K$. We consider the two cases of individual attacks (IA) and collective attacks (CA) by Eve, which result in two different key rates.   
In the case of individual attacks, Eve measures her ancillas before the classical postprocessing, so that at the beginning of it, Alice, Bob and Eve share a product probability distribution of classical symbols. In this case we have that 
\ba
\label{k-rate-ind-attack}
&K^{DR}=I(a,b)-I(a,E) \,, \\
&K^{RR}=I(a,b)-I(b,E)\,,
\ea
where $I(a,b)$ is the mutual information between the variables of Alice and Bob, and $I(a,E)$ ( $I(b,E)$) is the mutual information between Alice (Bob) and Eve.
For collective attacks, Eve keeps her ancillas in a quantum memory untill the end of the classical postprocessing, when she then performs an optimal collective measurement
\ba
\label{k-rate-coll-attack}
&K^{DR}=I(a,b)-S(a,E) \,, \\
&K^{RR}=I(a,b)-S(b,E)\,,
\ea
where now $S(a,E)$ ($S(b,E)$) is the Holevo bound on Eve's accessible information. 

It is possible to describe in a unifed way the different cases where homodyne or heterodyne measurements are performed. Including a beam splitter of transmissivity $T_A$ ($T_B$) at Alice's (Bob) side, where the ancillary mode is in the vacuum state of variance $V^{(0)}=1$, we easily find the modified variances and correlations
\begin{align}\label{va-vb}
\begin{split}
&V_{A^{T_A}}=T_A V_A+1-T_A  \,,\\
&V_{B^{T_B}}=T_B V_B+1-T_B  \,, \\
&\la\hat X_{A^{T_A}}\hat X_{B^{T_B}}\ra=\sqrt{T_AT_B}\,\la\hat X_{A}\hat X_{B}\ra \,.
\end{split}
\end{align}
Homodyne detection is recovered for $T_A=1$ ($T_B=1$), while heterodyning corresponds to $T_A=1/2$ ($T_B=1/2$). It has been shown that the four possible measurements correspond to the four different protocols mentioned in the introduction, in which Alice sends either squeezed states or coherent states, while Bob measures them with one of the two different techniques \cite{Gar}.     
The mutual information between Alice and Bob is given by the following equation
\begin{equation}\label{mut-ab}
I(a,b)=\frac{1}{2}\log\left(\frac{V_{A^{T_A}}}{V_{A^{T_A}|B^{T_B}}}\right) \,,
\end{equation} 
\\
and the conditional variances are given by
\begin{widetext}
\begin{align}\label{cond-var}
V_{A^{T_A}|B^{T_B}}=\la\hat X_{A^T}^2\ra-\frac{\la\hat X_{A^T}\hat X_{B^T}\ra}{\la\hat X_{B^T}^2\ra}=\begin{cases}
V_{A|B} & \text{if } T_A=T_B=1\,, \\
V_{A^M|B}=\frac{1}{2}(V_{A|B}+1) & \text{if }T_A=\frac{1}{2}\,,T_B=1 \,,\\
V_{A|B^M}=\frac{V_A}{V_B+1}(V_{B|A}+1)  & \text{if }T_A=1\,,T_B=\frac{1}{2} \,,\\
V_{A^M|B^M}=\frac{V_A}{2(V_B+1)}(V_{B|A}+1)+\frac{1}{2} & \text{if }T_A=\frac{1}{2}\,,T_B=\frac{1}{2}
\end{cases}
\end{align}
\end{widetext}
The letter $M$ means that the heterodyne measurement is performed. It has been proved that cloning attacks are optimal and saturate the bounds in $(\ref{k-rate-ind-attack})$ and $(\ref{k-rate-coll-attack})$ for the first three protocols, both for IA and CA. Therefore, we will consider only them in what follows. 

\subsection{Individual attacks}
In the case of individual attacks Eve's information for DR is easily computed as
\begin{equation}\label{mut-ae}
I(a,E)=\frac{1}{2}\log\left(\frac{V_{A^{T_A}}}{V_{A^{T_A}|E}}\right) \,,
\end{equation}  
where, according to whether Alice performs homodyne or heterodyne measurement we have
\begin{align}\label{cond-var-E}
V_{A^{T_A}|E}&=T_AV_{A|E}+1-T_A \nonumber\\
&=\begin{cases}
V_{A|E} & \text{if } T_A=1\,, \\
V_{A^M|E}=\frac{1}{2}(V_{A|E}+1) & \text{if }T_A=\frac{1}{2} \,.
\end{cases}
\end{align}
In the case of RR we need $V_{B^{T_B}|A^{T_A}}$ and $V_{B^{T_B}|E}$. It is enough to exchange $A$ with $B$ in the Eq. $(\ref{cond-var})$ and $(\ref{cond-var-E})$. 

\subsection{Collective attacks}\label{CA}
For collective attacks the Holevo bound in DR is given by
\begin{equation}
S(a,E)=S(E)-S(E|a) \,.
\end{equation}
Assuming that Eve holds the purification of the state allows us to write $S(E)=S(AB)$ and $S(E|a)=S(B|a)$, so that Eve's accessible information is a function of the entropic quantities of Alice and Bob only. 
In the EB scenario Alice prepares an EPR state (two modes squeezed vacuum) \cite{Gar,Wee} of zero mean 
\begin{equation}
d^{(in)}=(0,0)
\end{equation}
and covariance matrix
\begin{equation}\label{cov-in-AA'}
\sigma_{AA'}
=
\begin{bmatrix}
V\, I & \sqrt{V^2-1}\,\sigma_z\\ 
\sqrt{V^2-1}\, \sigma_z & V\, I 
\end{bmatrix} \,,
\end{equation}
with $\sigma_z$ being the Pauli matrix.
After one mode is sent to Bob through the noisy channel introduced by Eve $\mathcal{N}_{A'\to B}(\sigma_{AA'})=\sigma_{AB}$,
the covariance matrix of the output can be easily computed using $(\ref{Eve})$ 
\ba\label{cov-ine-AB}
\sigma_{AB}&=
\begin{bmatrix}
x\,I & z\,\sigma_z \\ 
z\,\sigma_z  & y\,I 
\end{bmatrix} \\
\\
&=\begin{bmatrix}
V\,I & \sqrt{T(V^2-1)}\,\sigma_z \\ 
\sqrt{T(V^2-1)}\,\sigma_z  & T(V+\chi)\,I 
\end{bmatrix} \,,
\ea
that is $x=V_A$ and $y=V_B$.
Given a two modes covariance matrix
\begin{equation}\label{cov-mat}
\sigma_{AB}
=\begin{bmatrix}
A & C\\ 
C^T & B 
\end{bmatrix}\,,
\end{equation} 
where $A$ and $B$ are $2\times2$ covariance matrices of the respective subsystems and $C$ is the correlation matrix, 
we have that
\begin{equation}\label{AB-entr}
S(AB)=g(\lambda_1)+g(\lambda_2)  \,,
\end{equation}
where $\lambda_{1,2}$ are the symplectic eigenvalues of $(\ref{cov-mat})$, given by 
\begin{align}
\begin{split}\label{simp-eigv}
\lambda_{1,2}^2&=\frac{1}{2}\left(\Delta\pm\sqrt{\Delta^2-4D^2}\right)  \,, \\
\Delta&=\det A+\det B+2 \det C  \,,\\
D&=\det\sigma_{AB} \,,  
\end{split}
\end{align}
and 
\begin{equation}
g(x)=\left(\frac{x+1}{2}\right)\log_2\left(\frac{x+1}{2}\right)-\left(\frac{x-1}{2}\right)\log_2\left(\frac{x-1}{2}\right) \,.
\end{equation}
Also, we have that 
\begin{equation}\label{con-entr}
S(B|a)=g(\lambda_3) \,,
\end{equation} 
$\lambda_3$ being the symplectic eigenvalue of Bob's covariance matrix $\sigma_{B|a}$, given Alice's measurement outcome, 
\begin{equation}\label{lam3}
\lambda_3=\sqrt{\det(\sigma_{B|a})} \,.
\end{equation}
When Alice performs a homodyne measurement on her mode, which gives $a$ as result, she projects Bob's mode on a state with mean 
\begin{align}
d_{B|a}=C^T(A_{11}^{-1}\Pi)\,d_A=\sqrt{1-1/V^2} \,(a,0) \,,
\end{align}
and covariance matrix 
\begin{align}
\sigma_{B|a}=B-C\,(A_{11}^{-1}\Pi)\,C^T \,,
\end{align}
where $\Pi=\text{diag}(1,0)$ and $d_A$ is Alice's measurement result. Note that this is a squeezed state displaced along the $x$ quadrature.
In the case of RR protocol we simply have to compute $S(A|b)$, which is done replacing $a$ with $b$ in the Eq. $(\ref{con-entr})$. 

The information leaked to Eve is harder to compute for the other two protocols.
In the case of coherent states and homodyne detection we attach the ancillary mode $C$ at Alice's side. The covariance matrix of the three modes reads  
\ba
\sigma_{ACB}&=
\begin{bmatrix}
(1/2)(x+1)\,I& (1/2)(1-x)\,I & \sqrt{1/2}\,z\,\sigma_z \\ 
(1/2)(1-x)\,I& (1/2)(x+1)\,I & -\sqrt{1/2}\,z\,\sigma_z \\ 
 \sqrt{1/2}\,z\,\sigma_z & -\sqrt{1/2}\,z\,\sigma_z  & y\,I  
\end{bmatrix}\,.
\ea
Now, the mutual information can be easily computed inserting in (\ref{mut-ab}) the variances from (\ref{va-vb}) and (\ref{cond-var}) corresponding to the cases $T_A=0.5$ and $T_B=1$. Also, while the Holevo bound of RR, $S(b,E)=S(E)-S(E|b)$, is exactly the same as that for squeezed states and homodyne detection - $(\ref{AB-entr})$ less $(\ref{con-entr})$ with $a\leftrightarrow b$ - the same problem for DR is more subtle because of the ancillary mode at Alice's side. The function $S(E)$ is still given by $(\ref{AB-entr})$ but, $S(E
|a)=S(CB|a)$ has now to be computed from the $6\times 6$ covariance matrix $\sigma_{BC|a}$. The result is \cite{Gar}
\b\label{E-acc-inf}
S(CB|a)=g(\lambda_4)+g(\lambda_5)\,,
\e
where the symplectic eigenvalues $\lambda_{4,5}$ are calculated again using the formula $(\ref{simp-eigv})$.
For squeezed states and heterodyne detection we have to take into account the ancillary mode $C$ at Bob's side. The covariance matrix is now 
\ba
\sigma_{ABC}&=
\begin{bmatrix}
x\,I & \sqrt{1/2}\,z\,\sigma_z &-\sqrt{1/2}\,z\,\sigma_z \\ 
 \sqrt{1/2}\,z\,\sigma_z& (1/2)(y+1)\,I &  (1/2)(1-y)I \\ 
-\sqrt{1/2}\,z\,\sigma_z & (1/2)(1-y)\,I  &  (1/2)(y+1)\,I
\end{bmatrix}\,,
\ea
The mutual information (\ref{mut-ab}) has to be computed for $T_A=1$ and $T_B=0.5$. 
In the case of DR Eve's information on Alice measurements is again the same as in DR for squeezed states and homodyne detection, $(\ref{AB-entr})$ less $(\ref{con-entr})$. 
To find Eve's accessible information $S(b,E)=S(E)-S(E|b)$ in RR, we need to find $\sigma_{AC|b}$ and $S(AC|b)$.
The result is that it is enough to change $x$ with $y$ in the eigenvalues $\lambda_{4,5}$. 

Now that we have all the standard machinery, we can move to the problem of how gravity affects the ability of Alice and Bob to perform QKD. To do so, we will assume that Alice and Bob start moving with acceleration $\mathcal{A}_I$ and $\mathcal{A}_{II}$, respectively.


\section{Effect of the Acceleration on Gaussian States}\label{Unruh channel}
We want to study the previously described protocols in the non-inertial scenario. The two reliable parties, Alice and Bob, are now allowed to accelerate independently one from the other. Let us consider two wave packets $\phi_{\Lambda}$ in the inertial frame,  where $\Lambda\in\left\{ I,II \right\}$, and two wave packets $\psi_{\Lambda}$ in the non inertial frame. These, are assumed to be localized and composed of positive frequencies centered around $\Omega_0$. In particular, the size $L$ of the accelerating modes have to be small enough to allow the attribution of a unique proper acceleration $\mathcal A_{\Lambda}$, $1/\mathcal{A}_{\Lambda}\ll L$. Also, they have to be far from the event horizon and satisfy $\Omega_0\gg 1/L$, which makes the contribution from negative frequencies negligible.  
A possible choice of the modes has been proposed in \cite{DDM,DDMB}. 
In the inertial frame they are  
\ba
&\phi_{\Lambda}(x,0)=\pm C\,e^{-2(\frac{x_0}{L}\log\frac{x}{x_0})^2}\sin\left(\sqrt{\Omega_0^2-m^2}(x-x_0)\right) , \\
&\partial_t\phi_{\Lambda}(x,0)=-i\,\Omega_0\,\phi_{\Lambda}(x,0)  \,,
\ea
for a wave packet centered around $x_0$. An accelerated reference frame in $1+1$ spacetime is properly described by the Rindler coordinates $(\eta,\chi)$
\begin{align}
t=\chi \sinh(a\eta) \,, \qquad\qquad x=\chi\cosh(a\eta) \,,
\end{align} 
where $a$ is a positive parameter and $(t,x)$ are the Minkowski coordinates. 
In Rindler coordinates, the mode functions are, accordingly, given by 
\small\ba
&\psi_{\Lambda}(\chi,0)=\pm C\,e^{-2(\frac{x_0}{L}\log(\frac{\chi}{x_0}))^2} \\
&\qquad\qquad\times \Im\left\{I_{\frac{-i\Omega_0}{\mathcal A}}(m|x_0|)I_{\frac{i\Omega_0}{\mathcal A}}(m|\chi|)\right\} \\
\\
&\partial_{\tau}\psi_{\Lambda}(\chi,0)=\mp\,i\,\Omega_0\,\psi_{\Lambda}(\chi,0)  \,,
\ea
where $I_{i\nu}(x)$ is the modified Bessel function of the first kind. 
Now, we need to properly describe how quantum states are transformed from one reference frame to the other. The most convincing and also flexible tool in this sense has been developed in \cite{Ahm}, where the authors were able to formulate the transformation in terms of a quantum channel acting on the covariance matrix of a two mode Gaussian states $\mathcal{N}(\sigma^{(f)})=\sigma^{(d)}$, 
\ba
\label{acc-cha}
X^{(d)}&=MX^{(f)}\,,  \\
\sigma^{(d)}&=M\sigma^{(f)}M^T+N  \,,
\ea
where $M$ and $N$ are $4\times 4$ real matrices and $N$ describes the noise present in the quantum channel.  
Here, $f$ and $d$ refer to the inertial modes $\phi_{\Lambda}$ and to the accelerated modes $\psi_{\Lambda}$, respectively. The vector of first moments is defined as $\vec{X}^{(i)}=\left\{ \hat q_I^{(i)}, \hat p_I^{(i)},\hat q_{II}^{(i)},\hat p_{II}^{(i)}\right\}$, $i\in \left\{f,d\right\}$. The Bogolubov transformations between the two set of modes, from one reference frame to the other, depend on the mode overlap 
\b
\hat d_{\Lambda}=(\psi_{\Lambda},\phi_{\Lambda})\hat f_{\Lambda} + (\psi_{\Lambda},\phi^*_{\Lambda})\hat f^{\dagger}_{\Lambda} \,.
\e
Defining 
\ba\label{Bog}
\alpha_{\Lambda}&=(\psi_{\Lambda},\phi_{\Lambda}) \,,\\
\beta_{\Lambda}&=-(\psi_{\Lambda},\phi^*_{\Lambda}) \,,
\ea
from  the first of the $(\ref{acc-cha})$ the $M$ matrix can be easily found to be  
\begin{equation}\small 
M=
\begin{bmatrix}
\Re(\alpha_I-\beta_I) &-\Im(\alpha_I+\beta_I) & 0&0 \\ 
\Im(\alpha_I-\beta_I) &\Re(\alpha_I+\beta_I) & 0&0 \\ 
0&0&\Re(\alpha_{II}-\beta_{II}) &-\Im(\alpha_{II}+\beta_{II})  \\ 
0&0&\Im(\alpha_{II}-\beta_{II}) &\Re(\alpha_{II}+\beta_{II})  
\end{bmatrix} \,.
\end{equation}
Regarding the noise matrix $N$, being more tricky to calculate, we refer the reader to \cite{Ahm}. Here, we mention only that for squeezed states, with sufficiently large squeezing parameters, as input to the channel, the $\beta_{\Lambda}$ coefficients can be neglected, being several orders of magnitude smaller than the $\alpha_{\Lambda}$. We assume to work inside this range, and the matrices simplify in     
\begin{equation}\label{M-matrix}
M=\alpha_I I\oplus \alpha_{II}I  \,,
\end{equation}
and 
\begin{equation}\label{N-matrix}
N=(1-\alpha_I^2) I\oplus (1-\alpha_{II}^2)I  \,.
\end{equation}

\begin{figure*}[]
\begin{minipage}[h]{0.3\textwidth}	
\includegraphics[width=2.0in]{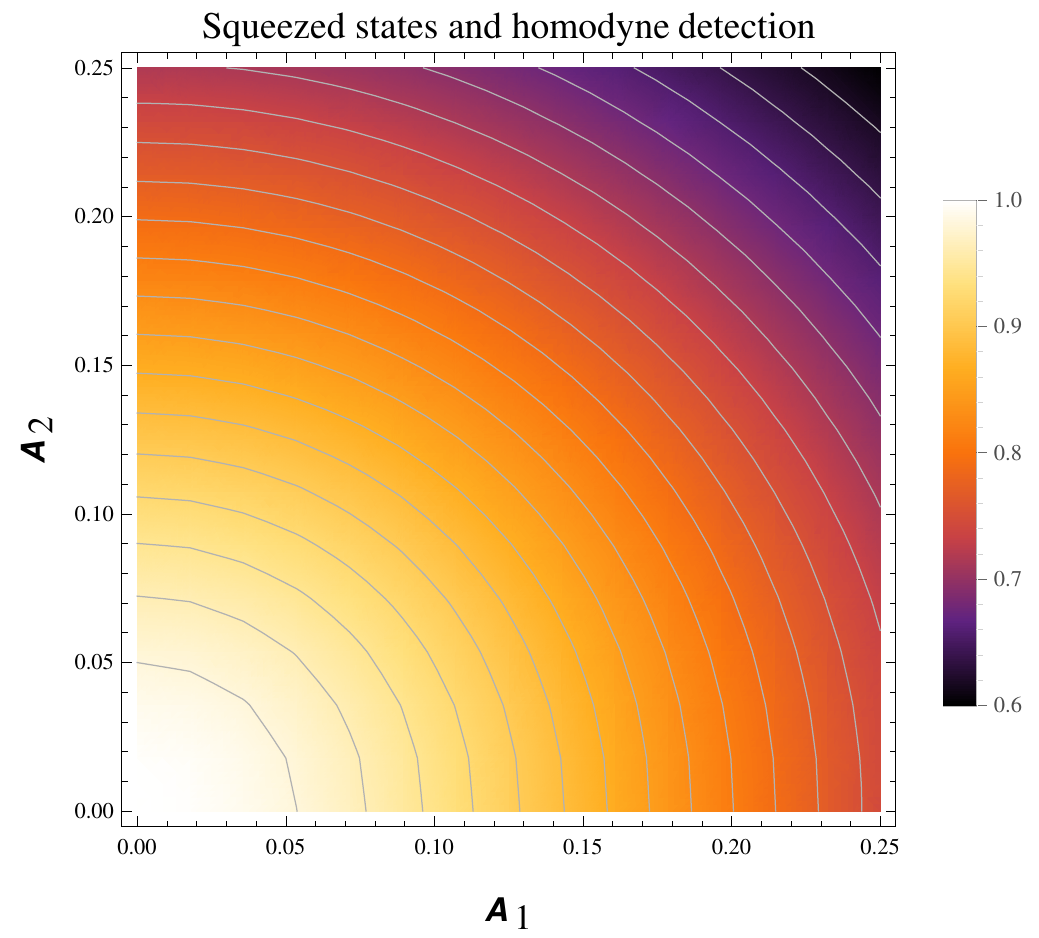}
\end{minipage}
\begin{minipage}[h]{0.3\textwidth}	
\includegraphics[width=2.0in]{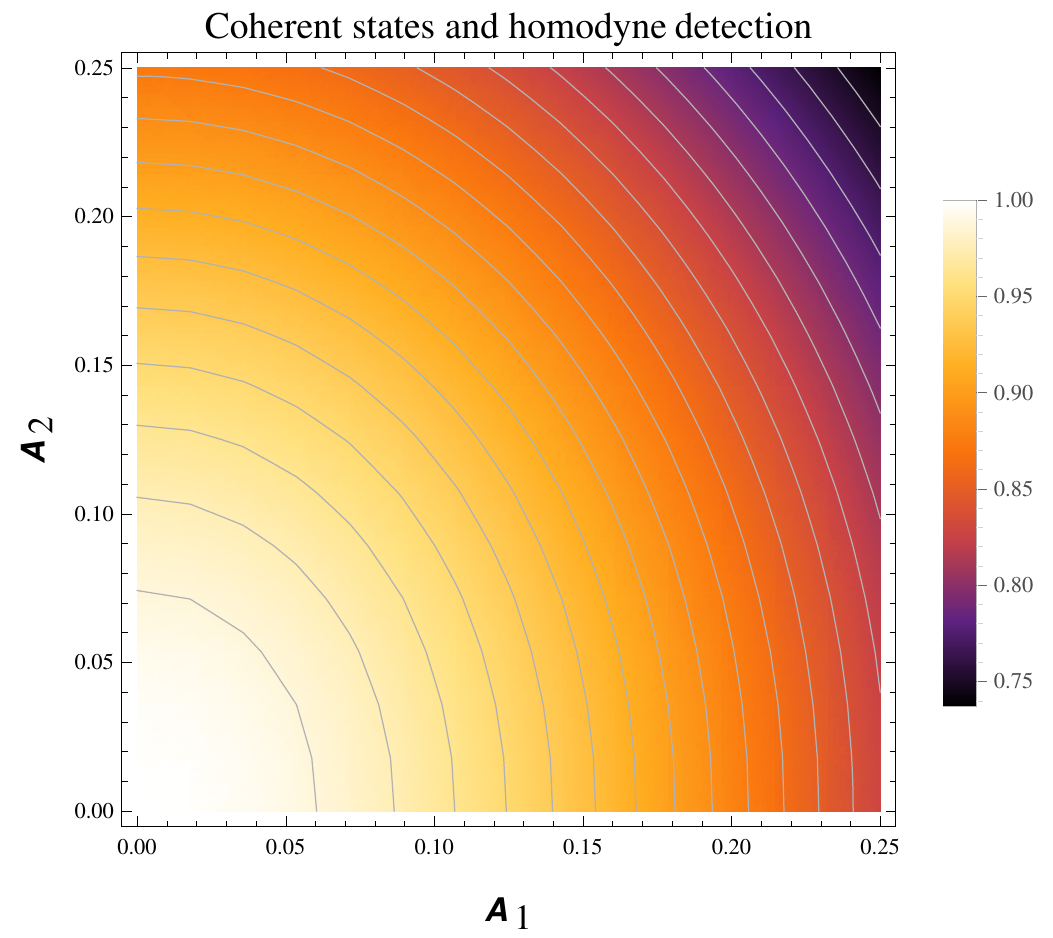}
\end{minipage}
\begin{minipage}[h]{0.3\textwidth}	
\includegraphics[width=2.0in]{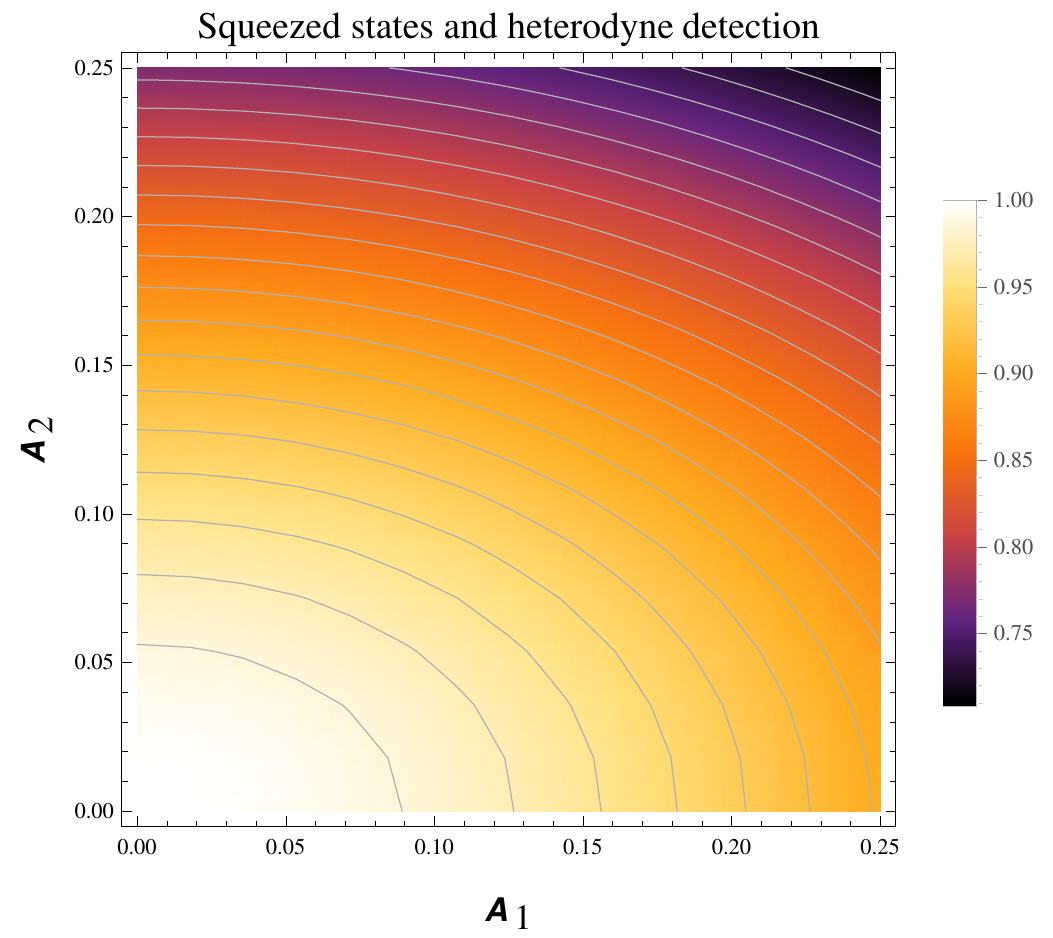}
\end{minipage}
\caption{Density plot of the ratio $K(\mathcal {A}_I,\mathcal {A}_{II})/K(0)$, for DR and IA, vs $\mathcal {A}_I$ and $\mathcal {A}_{II}$.
 The parameters are $\varepsilon=0.0$, $V=20$ and $l=0.5$.}
\label{fig1}
\end{figure*}

\begin{figure*}[]
\begin{minipage}[ptb]{0.3\textwidth}
\includegraphics[width=2.0in]{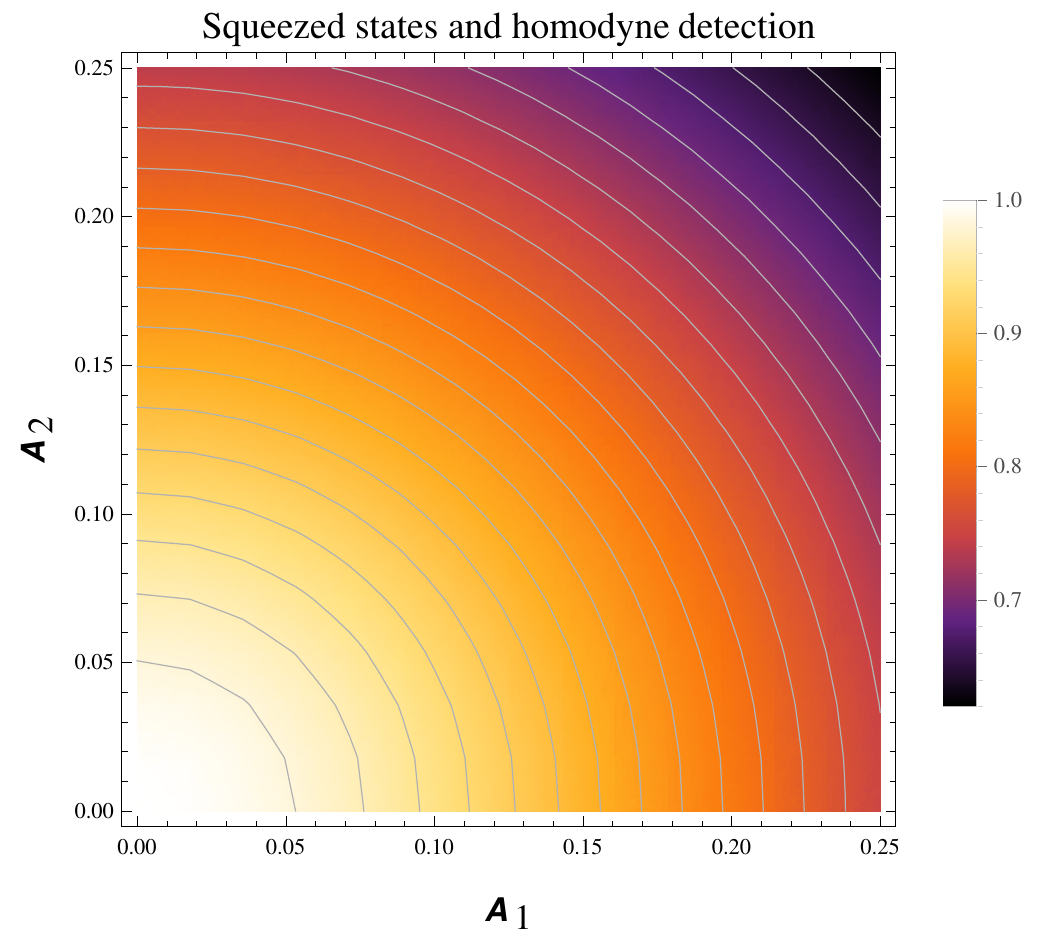}
\end{minipage}
\begin{minipage}[ptb]{0.3\textwidth}
\includegraphics[width=2.0in]{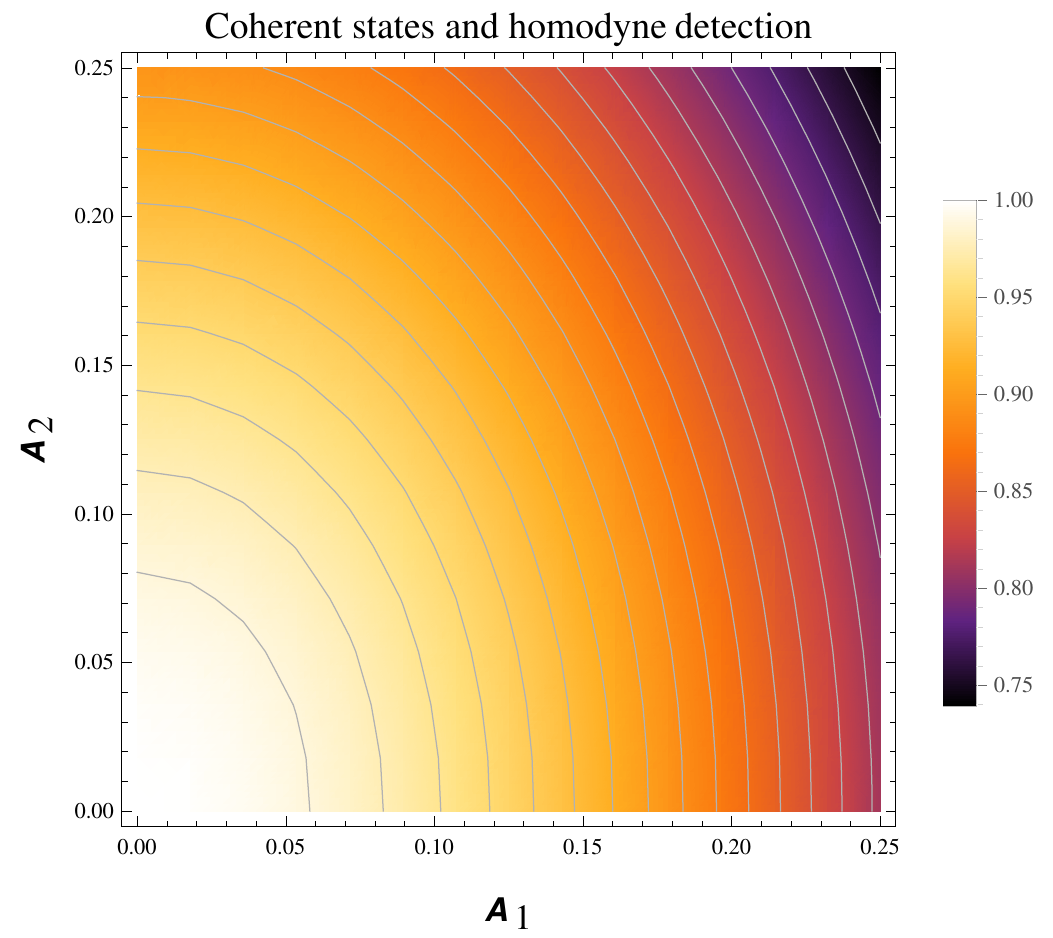}
\end{minipage}
\begin{minipage}[ptb]{0.3\textwidth}
\includegraphics[width=2.0in]{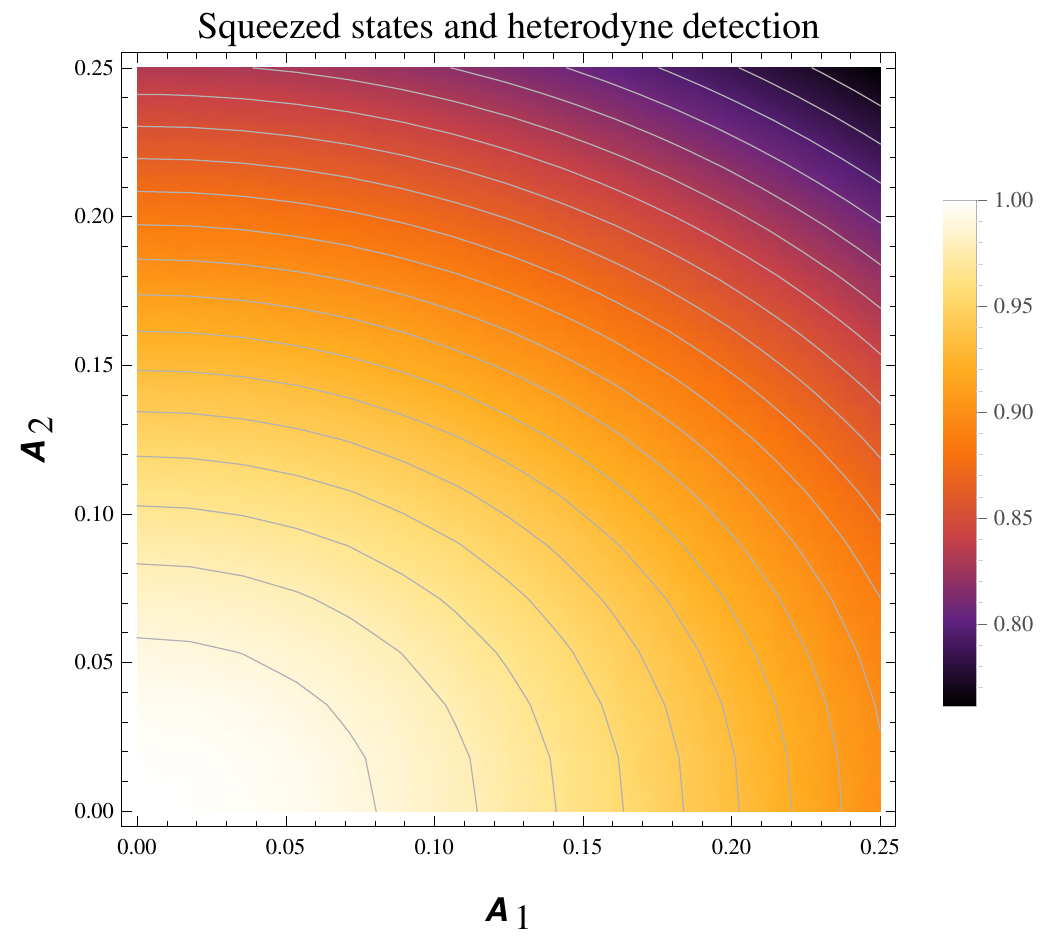}
\end{minipage}
\caption{Density plot of the ratio $K(\mathcal {A}_1,\mathcal {A}_2)/K(0)$, for RR and IA, vs $\mathcal {A}_I$ and $\mathcal {A}_{II}$.
 The parameters are $\varepsilon=0.0$, $V=20$ and $l=0.5$.}
\label{fig2}
\end{figure*}

\begin{figure*}[]
\begin{minipage}[ptb]{0.45\textwidth}
\includegraphics[width=3.0in]{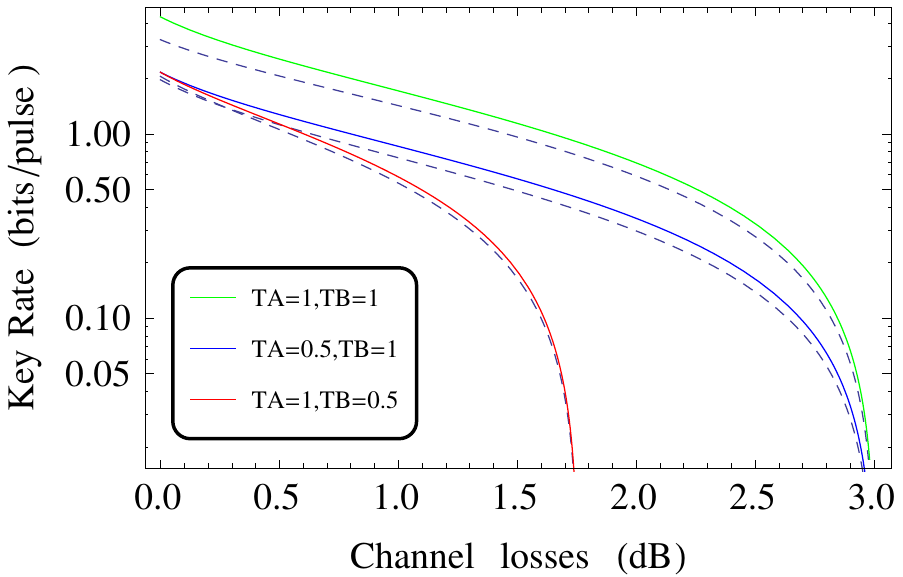}
\end{minipage}
\begin{minipage}[ptb]{0.45\textwidth}
\includegraphics[width=3.0in]{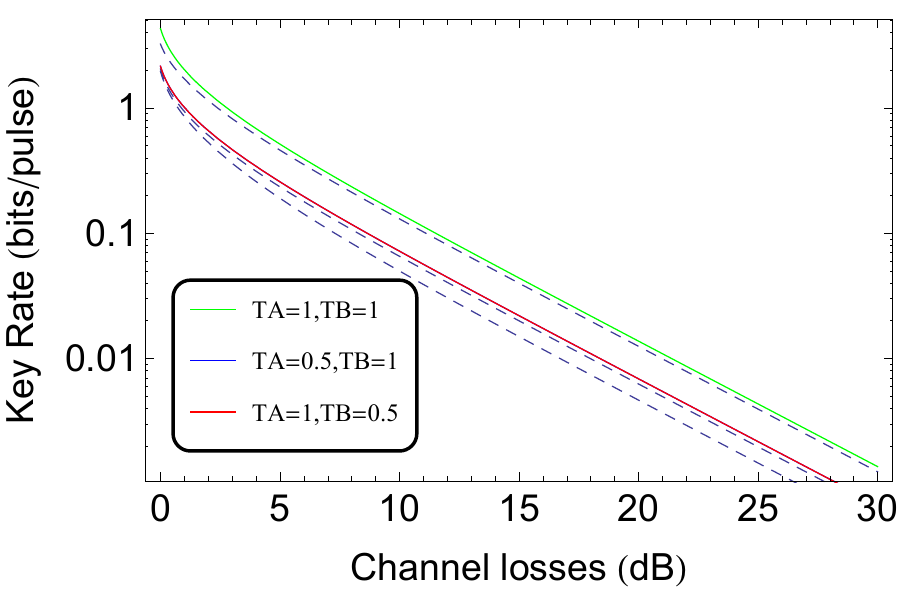}
\end{minipage}
\caption{$K(\mathcal {A})$ vs $l$, for IA and DR (left) and RR (right) respectively. Green line corresponds to squeezed states and homodyne detection, blue line to coherent states and homodyne detection, red line to squeezed states and heterodyne detection. The dashed lines are the respective key rate for non-inertial frames. 
The parameters are $\varepsilon=0.0$, $V=20$, $\mathcal{A}_1=0.2$ and $\mathcal{A}_2=0.0$.}
\label{fig3}
\end{figure*}

\begin{figure*}[]
\begin{minipage}[ptb]{0.45\textwidth}
\includegraphics[width=3.0in]{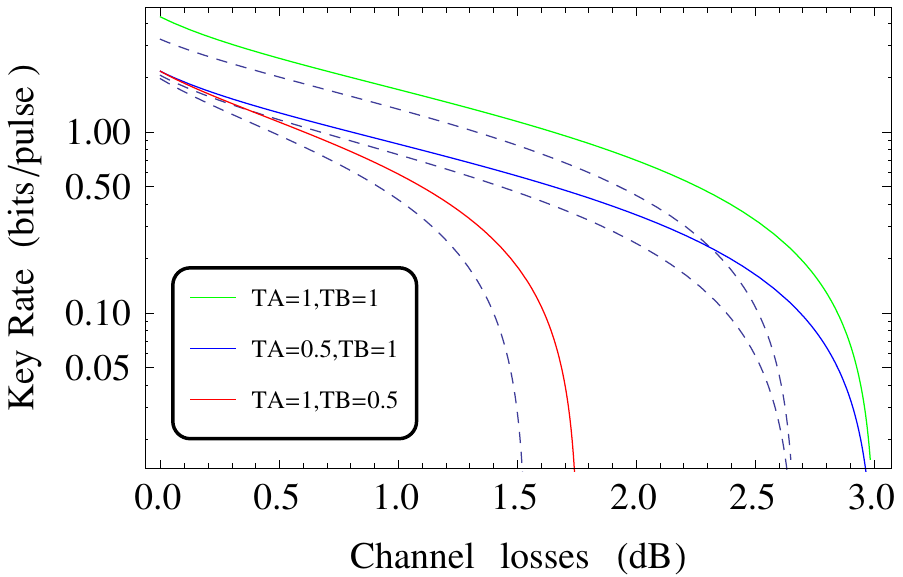}
\end{minipage}
\begin{minipage}[ptb]{0.45\textwidth}
\includegraphics[width=3.0in]{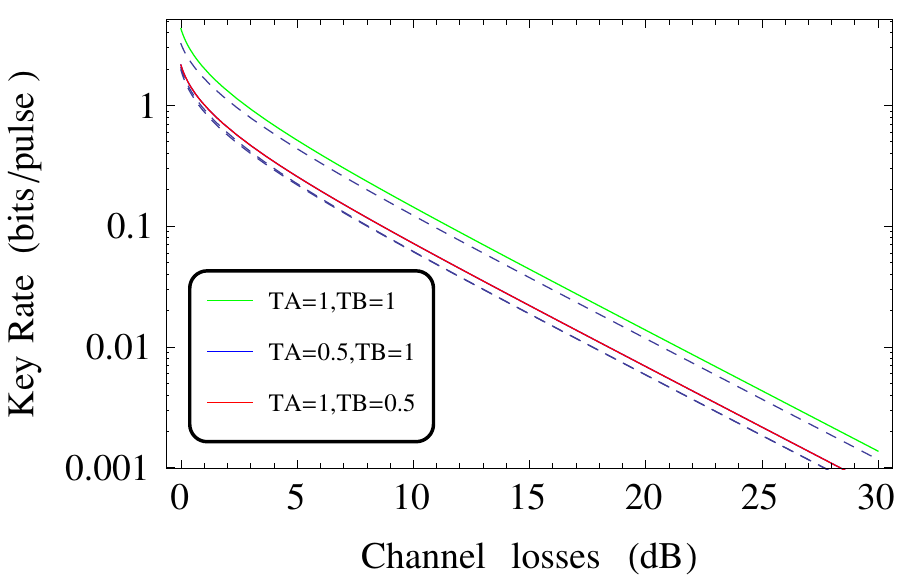}
\end{minipage}
\caption{$K(\mathcal {A})$ vs $l$, for IA and DR (left) and RR (right) respectively. Green line corresponds to squeezed states and homodyne detection, blue line to coherent states and homodyne detection, red line to squeezed states and heterodyne detection. The dashed lines are the respective key rate for non-inertial frames. 
The parameters are $\varepsilon=0.0$, $V=20$, $\mathcal{A}_1=0.0$ and $\mathcal{A}_2=0.2$.}
\label{fig4}
\end{figure*}


\section{Key rates in non inertial frames}\label{results} 
Applying to $(\ref{cov-ine-AB})$ the channel $(\ref{acc-cha})$, with $(\ref{M-matrix})$ and $(\ref{N-matrix})$, gives 
the following covariance matrix 
\begin{equation}\label{cov-mat-acc}
\sigma^{(d)}_{AB}=
\begin{bmatrix}
r\,I & t\,\sigma_z \\ 
t\,\sigma_z & s\,I   
\end{bmatrix} \,,
\end{equation}
where 
\ba\label{new-rst}
&r=\alpha_I^2\,x+(1-\alpha_I^2)  \,, \\
&s=\alpha_{II}^2\,y+(1-\alpha_{II}^2)    \,,\\
&t=\alpha_I\alpha_{II}\,z   \,.
\ea
As a consequence, the effect of gravity on the protocols can be accounted for simply exchanging the quantities $x$, $y$ and $z$ with $r$, $s$ and $z$. This holds for all the cases previosly discussed, both in IA and CA regime. The quantities $\alpha_I$ and $\alpha_{II}$ are the Bogoliubov coefficients defined in $(\ref{Bog})$ and computed in the paper \cite{Ahm}. An important detail has to be underlined: we are here assuming that only the modes $A$ and $B$ are accelerating, while the ancilla $C$ is inertial.
The results are shown in the two following sections.
\begin{figure}[ptb]
\includegraphics[width=3.0in]{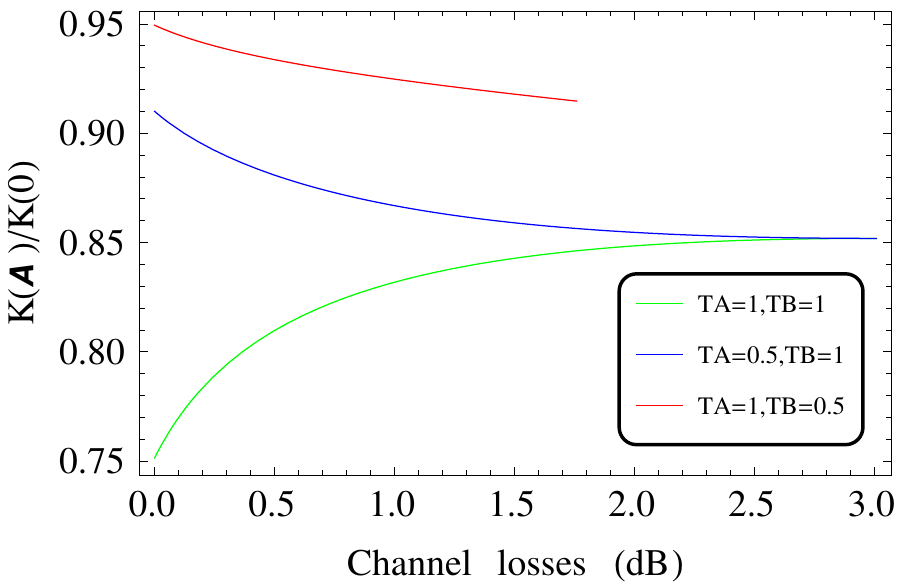}
\caption{Plots of the ratio $K(\mathcal {A})/K(0)$ vs $l$, for IA and DR. Green line corresponds to squeezed states and homodyne detection, blue line to coherent states and homodyne detection, red line to squeezed states and heterodyne detection.
The parameters are $\varepsilon=0.0$, $V=20$, $\mathcal{A}_1=0.2$ and $\mathcal{A}_2=0.0$.}
\label{fig5}
\end{figure}

\begin{figure}[ptb]
\includegraphics[width=3.0in]{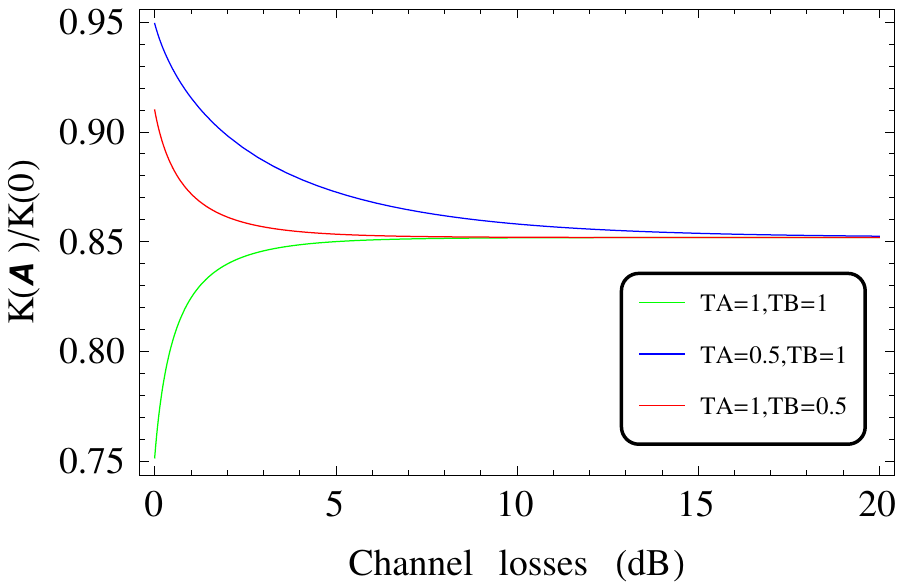}
\caption{Plots of the ratio $K(\mathcal {A})/K(0)$ vs $l$, for IA and RR. Green line corresponds to squeezed states and homodyne detection, blue line to coherent states and homodyne detection, red line to squeezed states and heterodyne detection. The parameters are $\varepsilon=0.0$, $V=20$, $\mathcal{A}_1=0.0$ and $\mathcal{A}_2=0.2$.}
\label{fig6}
\end{figure}

\begin{figure*}[]
\begin{minipage}[h]{0.3\textwidth}	
\includegraphics[width=2.0in]{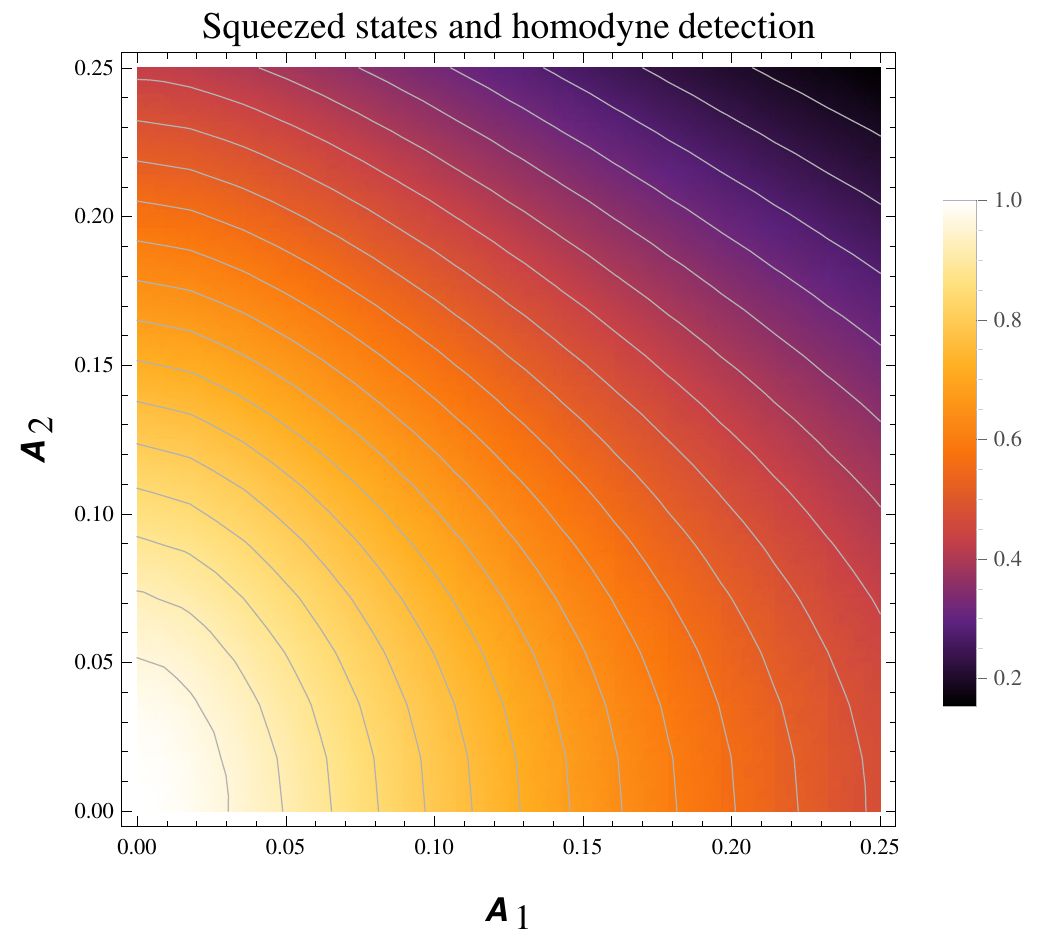}
\end{minipage}
\begin{minipage}[h]{0.3\textwidth}	
\includegraphics[width=2.0in]{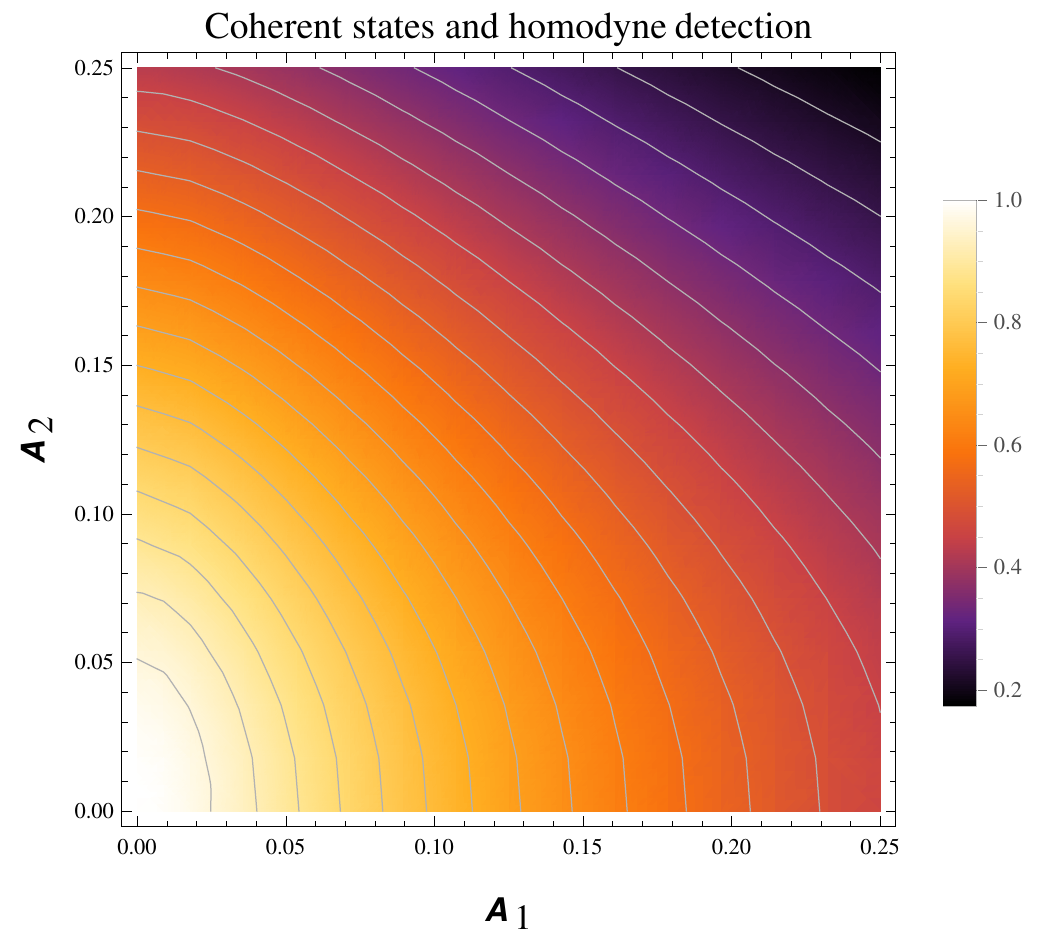}
\end{minipage}
\begin{minipage}[h]{0.3\textwidth}	
\includegraphics[width=2.0in]{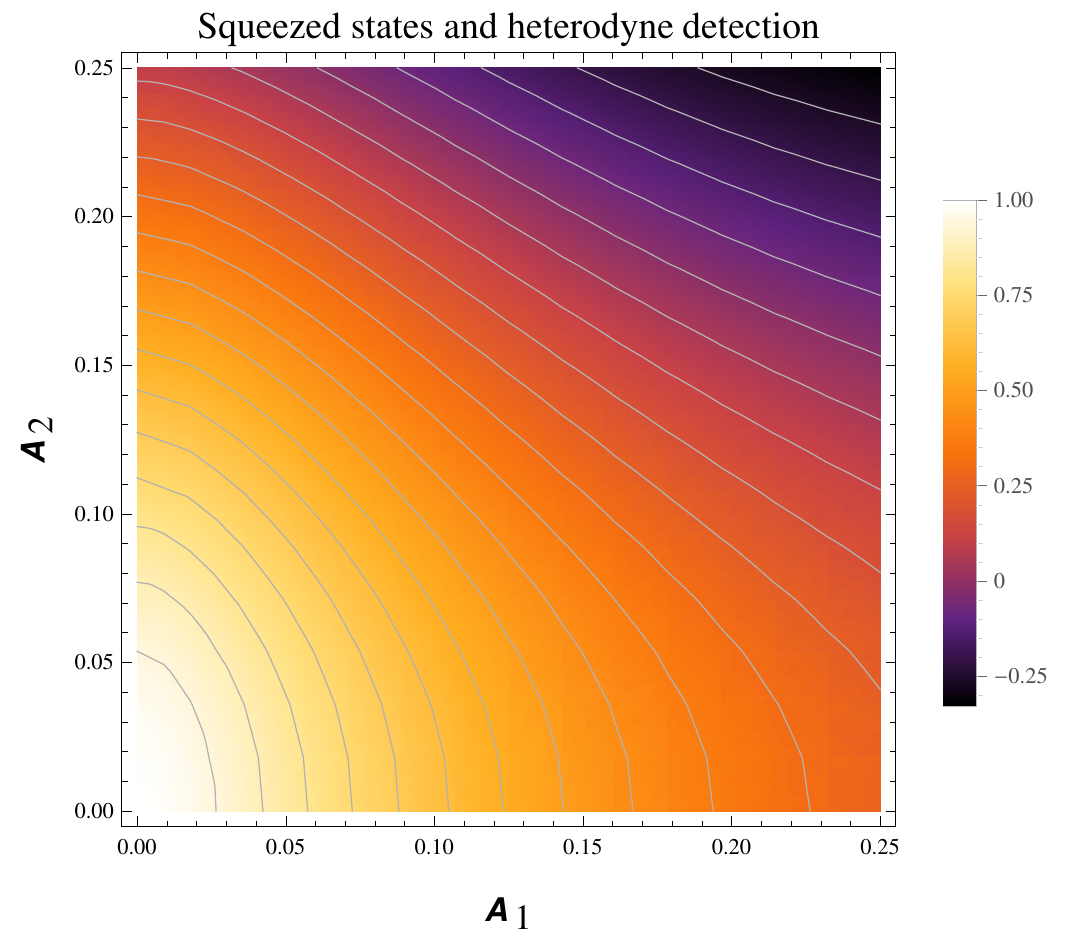}
\end{minipage}
\caption{Density plot of the ratio $K(\mathcal {A}_I,\mathcal {A}_{II})/K(0)$, for DR and CA, vs $\mathcal {A}_I$ and $\mathcal {A}_{II}$.
 The parameters are $\varepsilon=0.0$, $V=20$ and $l=0.5$.}
\label{fig7}
\end{figure*}

\begin{figure*}[]
\begin{minipage}[h]{0.3\textwidth}
\includegraphics[width=2.0in]{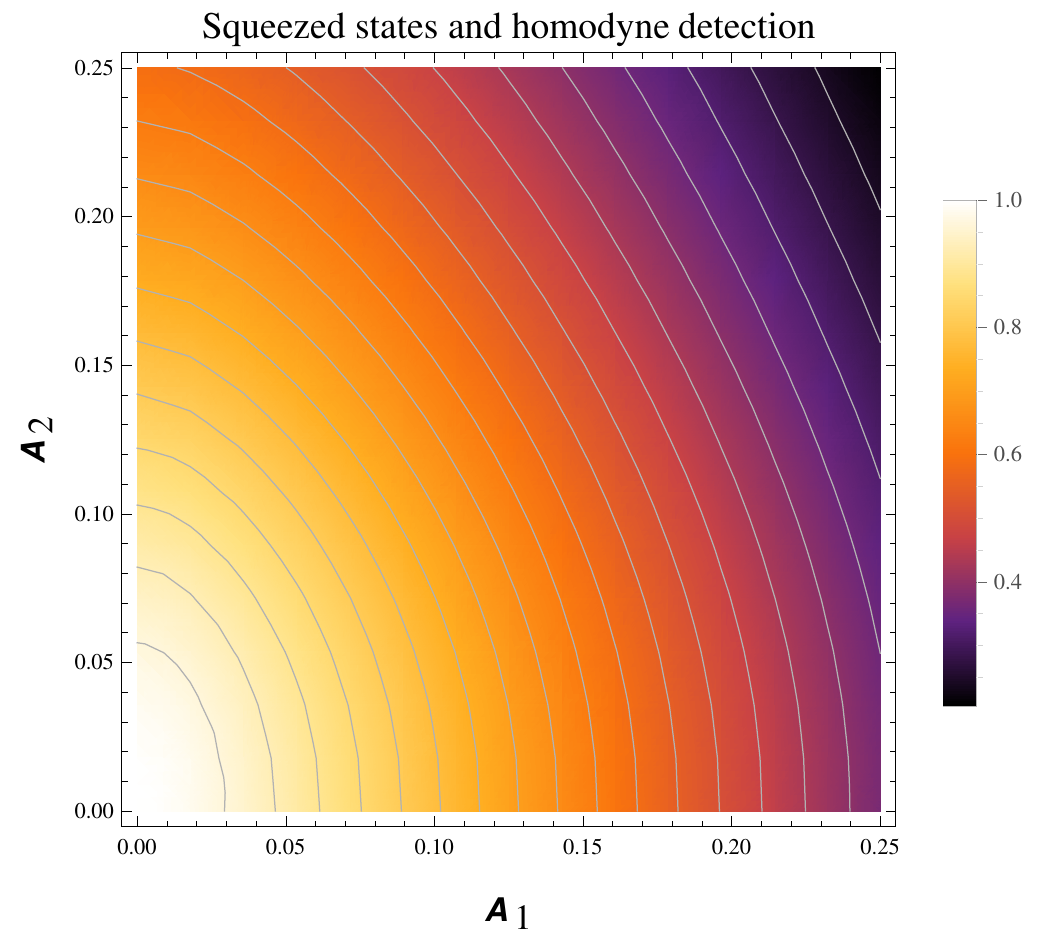}
\end{minipage}\hfil
\begin{minipage}[h]{0.3\textwidth}
\includegraphics[width=2.0in]{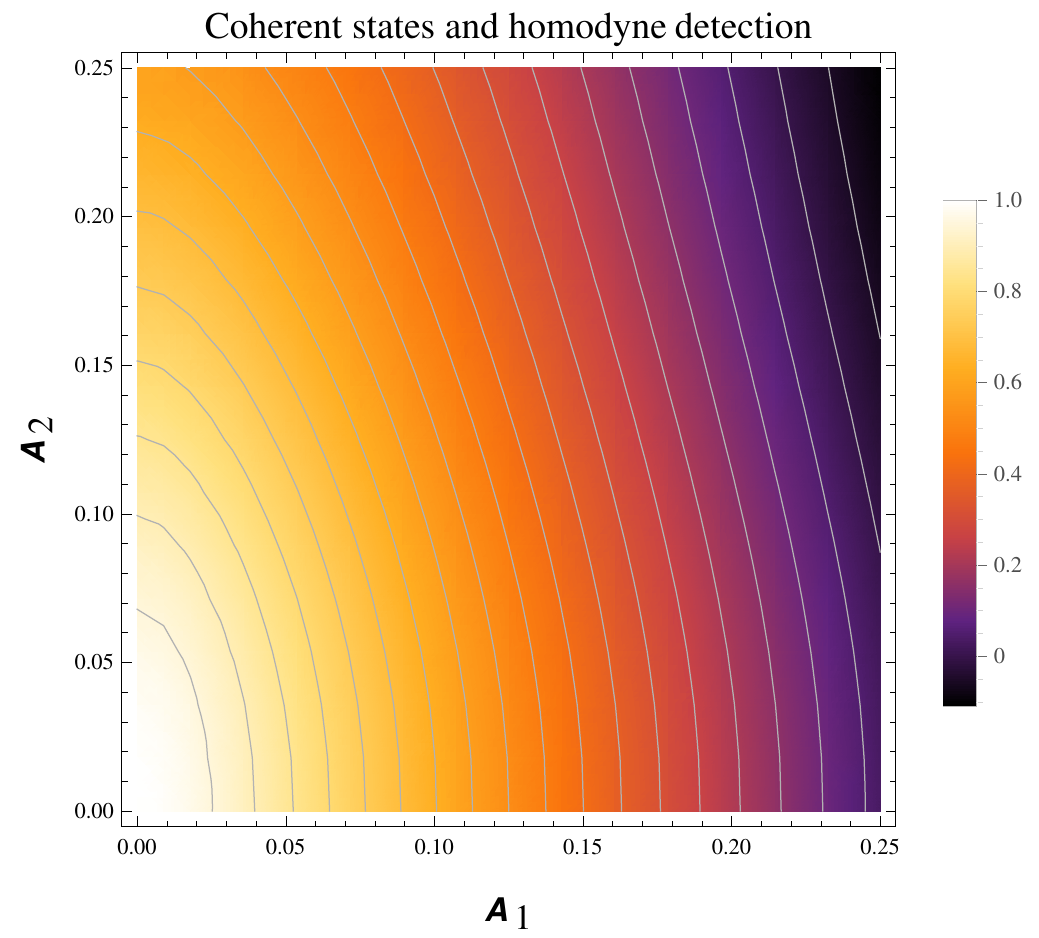}
\end{minipage}
\begin{minipage}[h]{0.3\textwidth}
\includegraphics[width=2.0in]{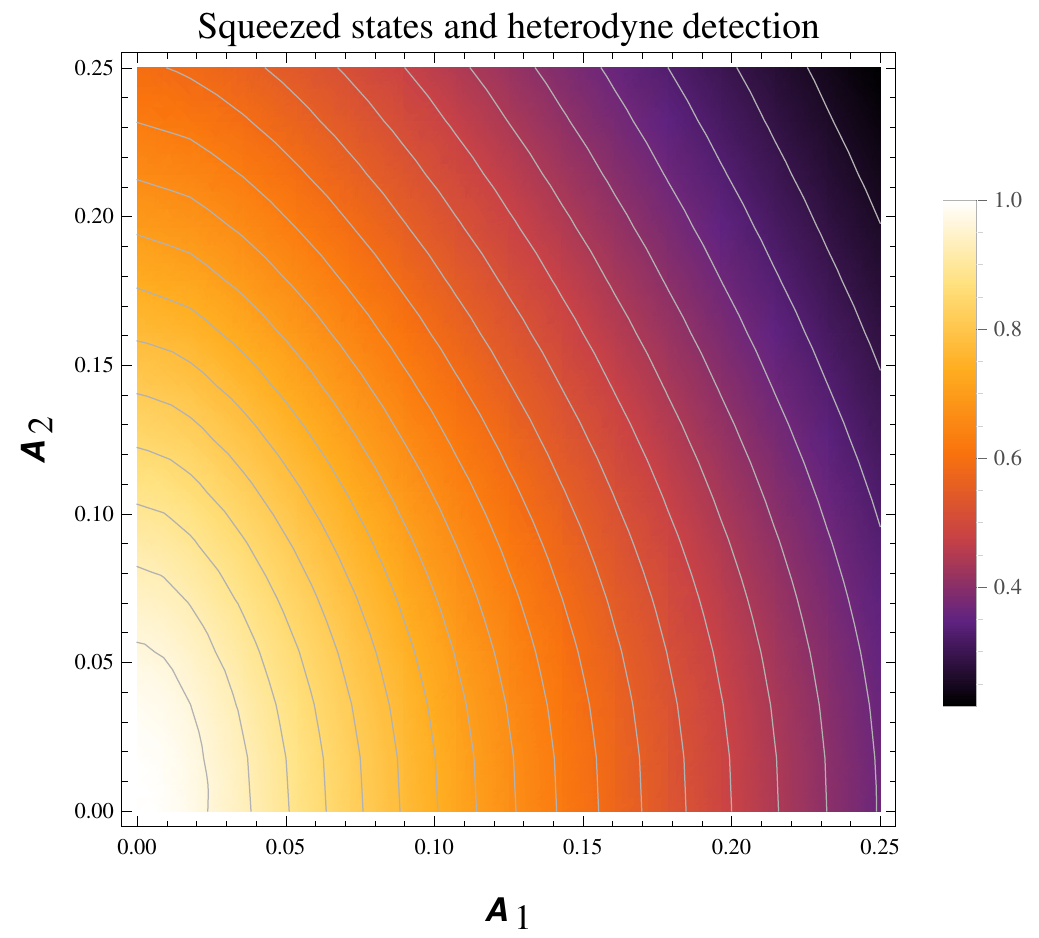}
\end{minipage}
\caption{Density plot of the ratio $K(\mathcal {A}_1,\mathcal {A}_2)/K(0)$, for RR and CA, vs $\mathcal {A}_I$ and $\mathcal {A}_{II}$.
 The parameters are $\varepsilon=0.0$, $V=20$ and $l=0.5$.}
\label{fig8}
\end{figure*}

\subsection{Individual attacks}
The key rate of individual attacks, for DR and RR, are given by the difference between the equations $(\ref{mut-ab})$ and $(\ref{mut-ae})$. We have
\begin{align}
K^{DR}&=\frac{1}{2}\log_2\frac{V_{A|E}}{V_{A|B}}  \,,\\
K^{RR}&=\frac{1}{2}\log_2\frac{V_{B|E}}{V_{B|A}}  \,.
\end{align}
The conditional variances for the accelerated modes are obtained introducing the quantities $(\ref{new-rst})$ into $(\ref{cond-var})$ and $(\ref{cond-var-E})$; they can be written in a compact form as
\b
V^{(d)}_{A^{T_A}|B^{T_B}}=\frac{\alpha_I^2\,V_{A^T}}{V_{B^T}+\frac{1}{\alpha_{II}^2}-1}\left( 
V^{(f)}_{B^{T_B}|A^{T_A}}+\frac{1}{\alpha_{II}^2}-1\right)+1-\alpha_I^2  \,,
\e
and
\b
V^{(d)}_{A^{T_A}|E}=\alpha_I^2\,V^{(f)}_{A^{T_A}|E}+1-\alpha_1^2 \,. 
\e
where $V^{(f)}_{B^{T_B}|A^{T_A}}$ and $V^{(f)}_{A^{T_A}|E}$ are defined in the equations (\ref{cond-var}) and (\ref{cond-var-E}).
The variances of Bob, given Alice or Eve measurements, can be calculated simply switching $A$ with $B$ and $I$ with $II$.

The results, in the case of individual attacks, are summarized in the plots. 
Fig. $(\ref{fig1})$ and  $(\ref{fig2})$ show the key rates $(\ref{k-rate-ind-attack})$ normalized to one (which corresponds to the inertial case) as a function of Alice and Bob's acceleration, for DR and RR respectively. Not surprisingly, the acceleration erodes the ability of the two parties to perform QKD in all the protocols.   
We can note that the degradation effect is almost symmetric respect to Alice and Bob's acceleration only for squeezed states and homodyne detection, both in DR and in RR. The other protocols are not symmetric and Alice and Bob's acceleration degrades the key with a different rate.
Fig. (\ref{fig3}) shows the key rates of the different protocols as a function of the loss $l$ \cite{loss} for DR and RR respectively, and with fixed variance of the initial shared entangled state $V$ and fixed noise $\varepsilon$. Next to each curve is represented the key rate of the relative protocol when Alice is undergoing a given acceleration and Bob is inertial (dashed lines), $\mathcal{A}_I\neq 0$ and $\mathcal{A}_{II}=0$. We note that the rate is generally lower in the non-inertial case, but, 
interestingly, in DR the maximal communication length is not affected. 
This interesting feature is lost when also Bob is accelerating and the transmission length becomes shorter. In RR the length is confirmed to be virtually infinite, but the rate is generally lower when gravity plays a role. Fig. (\ref{fig4}) shows the key rate when $\mathcal{A}_I=0$ and $\mathcal{A}_{II}\neq0$. 

The Fig. (\ref{fig5}) and (\ref{fig6}) represent the ratio between the non-inertial and the inertial rates as functions of the losses, for the various protocols. They also show some interesting features. When the reference party (Alice) is accelerating all of them converge to various constant values which shift to lower values as the acceleration increases. In particular, DR is shown in the Fig. (\ref{fig5}) for the case $\mathcal{A}_I\neq 0$ and $\mathcal{A}_{II}=0.0$. We observe that the protocols with coherent states and homodyne detection converges to the same value as the protocol with squeezed states and homodyne measurement. If  Bob is also accelerating the plots do not converge. A general features of the blue and red lines is that they are monotonically decreasing and when $\mathcal{A}_{II}=0$ they converge to some constant value. The green curve in DR is monotonically growing only in the case $\mathcal{A}_{II}=0$ and, as soon as Bob accelerates, it becomes decreasing at some values of the loss. In RR, when $\mathcal{A}_I=0$ and $\mathcal{A}_{II}\neq 0$, it is interesting to note that all the protocols converge to the same value, which changes with the acceleration, see Fig. (\ref{fig6}). When Alice is also accelerating the protocol with coherent states and homodyne detection moves away from this common value but the monotonic behaviour of the ratios does not change.

\begin{figure*}
\begin{minipage}[h]{0.45\textwidth}
\includegraphics[width=3.0in]{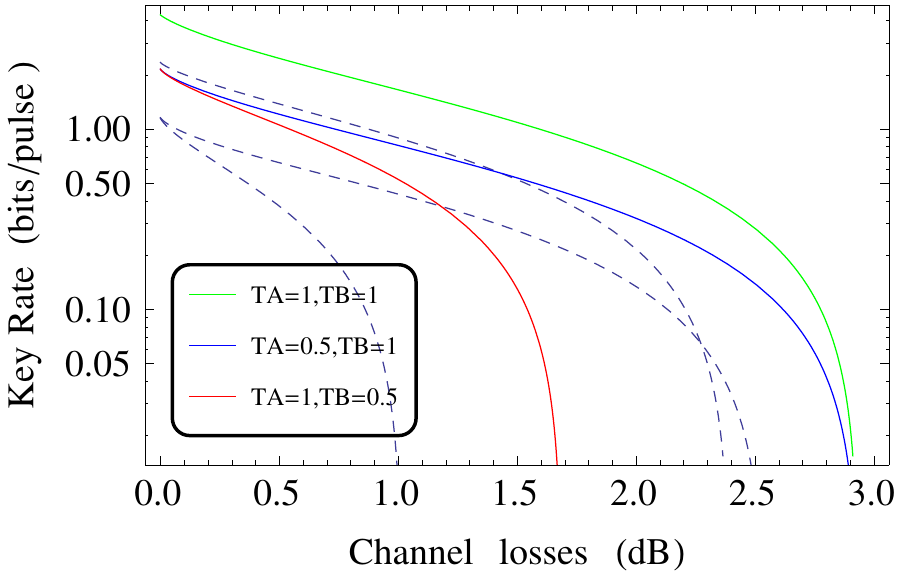}
\end{minipage}
\begin{minipage}[h]{0.45\textwidth}
\includegraphics[width=3.0in]{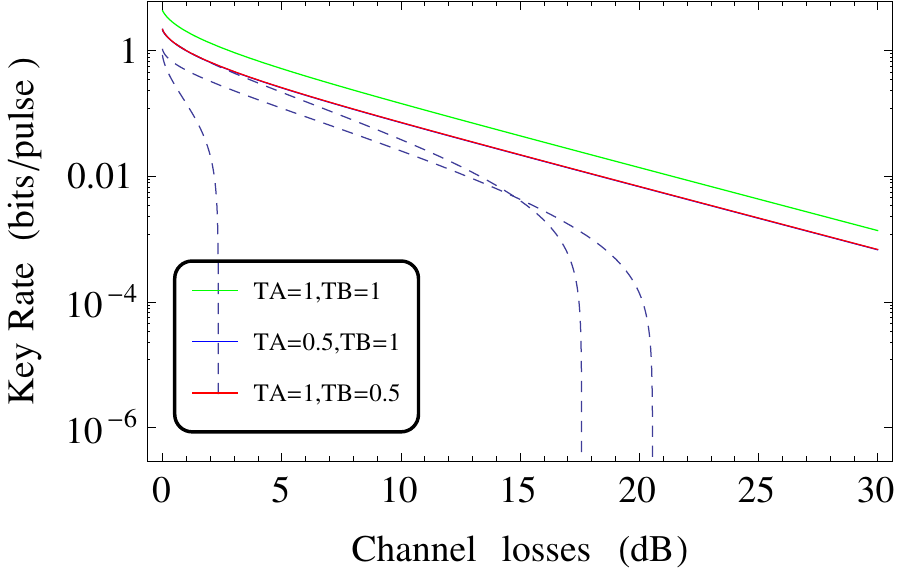}
\end{minipage}
\caption{$K(\mathcal {A})$ vs $l$, for CA and DR(left) and RR (right). Green line corresponds to squeezed states and homodyne detection, blue line to coherent states and homodyne detection, red line to squeezed states and heterodyne detection. The dashed lines are the respective key rate in non-inertial frames. 
The parameters are $\varepsilon=0.0$, $V=20$, $\mathcal{A}_1=0.2$ and $\mathcal{A}_2=0.0$.}
\label{fig9}
\end{figure*}

\begin{figure*}
\begin{minipage}[h]{0.45\textwidth}
\includegraphics[width=3.0in]{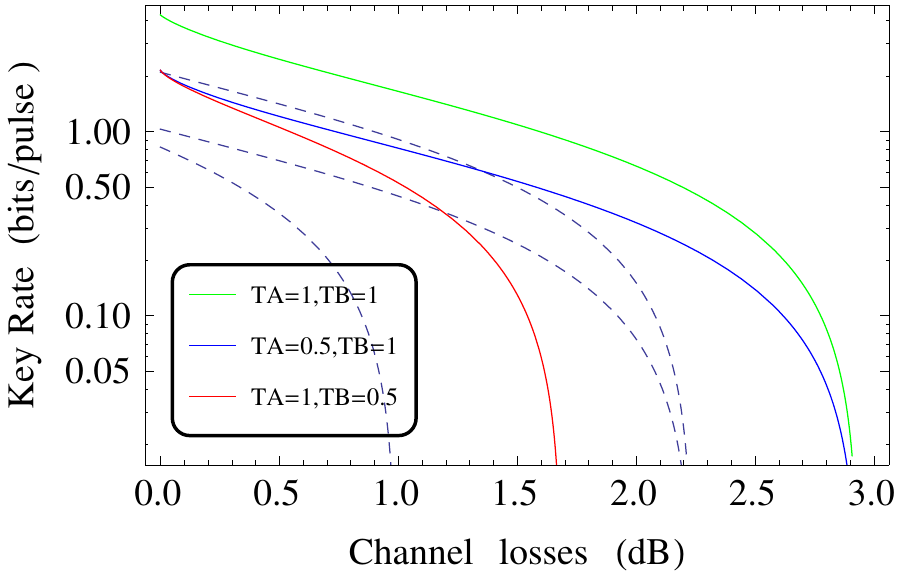}
\end{minipage}
\begin{minipage}[h]{0.45\textwidth}
\includegraphics[width=3.0in]{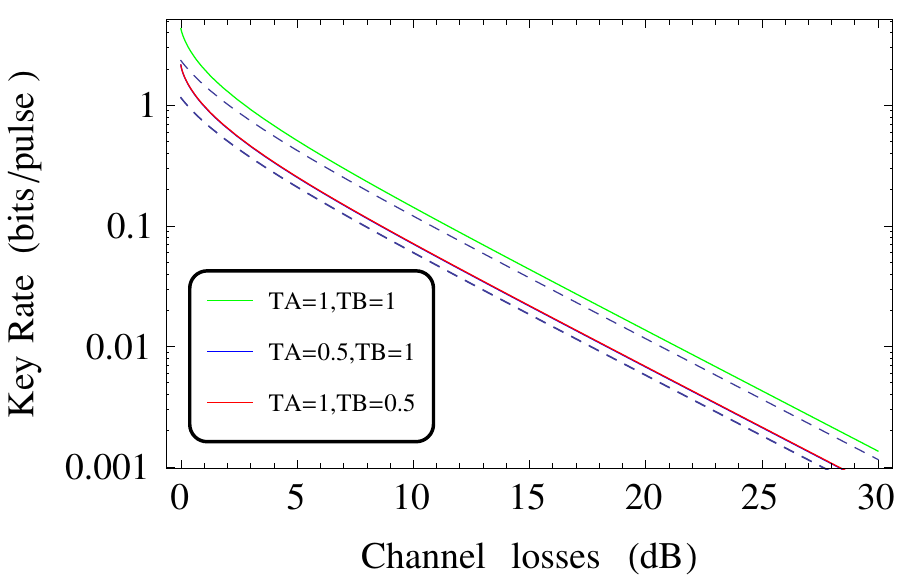}
\end{minipage}
\caption{$K(\mathcal {A})$ vs $l$, for CA and DR(left) and RR (right). Green line corresponds to squeezed states and homodyne detection, blue line to coherent states and homodyne detection, red line to squeezed states and heterodyne detection. The dashed lines are the respective key rate in non-inertial frames. 
The parameters are $\varepsilon=0.0$, $V=20$, $\mathcal{A}_1=0.0$ and $\mathcal{A}_2=0.2$.}
\label{fig10}
\end{figure*}

\subsection{Collective attacks}
\begin{figure}[h]
\includegraphics[width=3.0in]{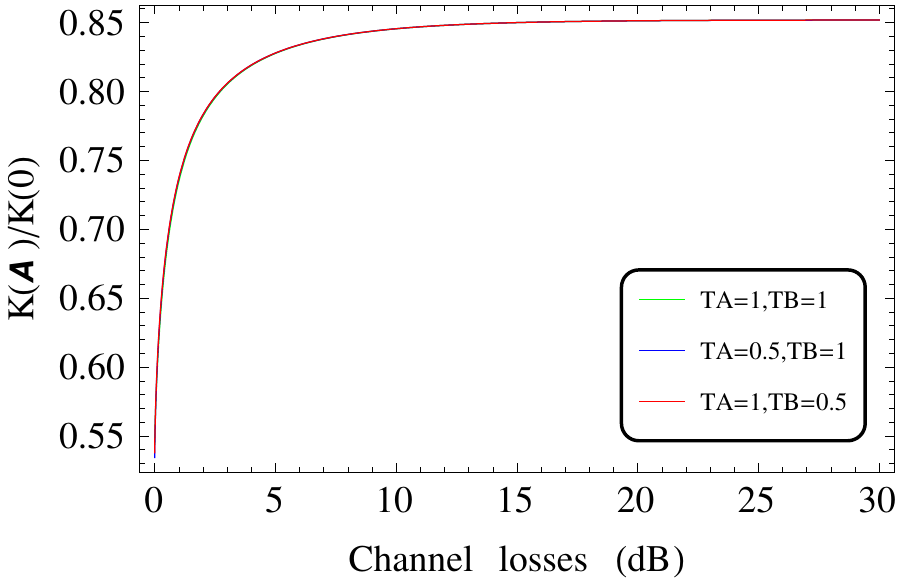}
\caption{Plots of the ratio $K(\mathcal {A})/K(0)$ vs $l$, for collective attacks and RR. The parameters are $\varepsilon=0.0$, $V=20$, $\mathcal{A}_1=0.0$ and $\mathcal{A}_2=0.2$.}
\label{fig11}
\end{figure}

For collective attacks we need the mutual information between Alice and Bob data 
\begin{equation}\label{AB-mutual-inf}
I(a,b)=\frac{1}{2}\log\left(\frac{V_{A^{T_A}}^{(d)}}{V^{(d)}_{A^{T_A}|B^{T_B}}}\right) \,.
\end{equation}   
Eve's accessible information, for squeezed states and homodyne detection, for DR and RR respectively, is
\ba\label{E-acc-inf}
S(a,E)&=g(\lambda_1)+g(\lambda_2)-g(\lambda^{(DR)}_3) \,,\\
S(b,E)&=g(\lambda_1)+g(\lambda_2)-g(\lambda^{(RR)}_3) \,,
\ea  
where now $\lambda_{1,2}$ are the symplectic eigenvalues of $(\ref{cov-mat-acc})$ 
\begin{align}\label{eigen}
\lambda_{1,2}&=\sqrt{\frac{1}{2}\left(\Delta\pm\sqrt{\Delta^2-4D^2}\right)}  \,, \nonumber\\
\Delta&=r^2+s^2-2t^2  \,,\\
D&=rs-t^2  \,. \nonumber
\end{align}
Furthermore,
$\lambda^{(DR)}_3$ and $\lambda^{(RR)}_3$ are the symplectic eigenvalues of $\sigma^{(d)}_{B|a}$ and $\sigma^{(d)}_{A|b}$ respectively, namely
\ba\label{E-acc-inf}
\lambda^{(DR)}_3=s(s-t^2/r)  \,,\\
\lambda^{(RR)}_3=r(r-t^2/s)  \,.
\ea   
In the same way we can write down the expressions for the other protocols, defined in section $(\ref{CA})$, making the substitution $x\to r$, $y\to s$ and $z\to t$. Coherent states and homodyne detection allow Eve to access the following amount of information,
\ba\label{E-acc-inf}
S(a,E)&=g(\lambda_1)+g(\lambda_2)-g(\lambda^{(DR)}_3) \,,\\
S(b,E)&=g(\lambda_1)+g(\lambda_2)-g(\lambda_4)-g(\lambda_5) \,,
\ea 
where the symplectic eigenvalues are calculated to be
\b\label{ABC-eig}
\lambda^2_{4,5}=\frac{1}{2}\left(A\pm\sqrt{A^2-4B}\right) \,,
\e
with
\begin{align}\label{AB}
\begin{split}
A &= \frac{1}{r+1}[r+sD+\Delta]\,, \\ 
B &= \frac{D}{r+1}[s+D] \,.
\end{split}
\end{align}
Analogously, for squeezed states and heterodyne detection we have
\ba\label{E-acc-inf}
S(a,E)&=g(\lambda_1)+g(\lambda_2)-g(\lambda_4)-g(\lambda_5) \,,\\
S(b,E)&=g(\lambda_1)+g(\lambda_2)-g(\lambda^{(RR)}_3) \,,
\ea 
where $\lambda_{4,5}$ are still given by the expression $(\ref{AB})$, but now with
\begin{align}
\begin{split}
A &= \frac{1}{s+1}[s+rD+\Delta]\,, \\ 
B &= \frac{D}{s+1}[r+D] \,.
\end{split}
\end{align}
Fig. $(\ref{fig7})$ and $(\ref{fig8})$ show the key rates $(\ref{k-rate-coll-attack})$ for CA, normalized to one, as a function of the accelerations of Alice and Bob, for DR and RR respectively. We can already notice a difference compared to IA: the degradation effect is not symmetric in the two parties in any of the protocols. Also, the behavior seems to be less regular. An other important difference is revealed in Fig. $(\ref{fig9})$, where the three protocols of DR and RR with $\mathcal{A}_1\neq 0$ and $\mathcal{A}_2=0$ are plotted. Unlike in IA regime, for CA the transmission length is shortened in any case by the acceleration of either Alice or Bob. Moreover, the key rates of all the protocols become practically zero at sufficiently high loss whenever $\mathcal{A}_1\neq 0$ even in RR, whereas they are always positive for IA. In Fig. $(\ref{fig10})$ we see that in the case $\mathcal{A}_1=0$ and $\mathcal{A}_2\neq0$, the protocols performance for RR are not strongly affected. 
The plot of the ratios vs losses in RR also reveals some interesting features. In particular, Fig. $(\ref{fig11})$ shows that the ratios for the three different protocols completely overlap when only Bob accelerates, $\mathcal{A}_1= 0$ and $\mathcal{A}_2\neq0$. The plots are growing monotonic functions and the convergence value increases when Bob's acceleration decreases. We did not find some interesting regularity for DR and CA protocols and, therefore, we do not show their plots here.  

\section{Concluding Remarks}\label{conclusion}

We have provided a very general method to consider the effects of gravity on several CV-QKD protocols. As expected, the performance is mostly deteriorated when the motion is non-inertial. Nevertheless, few results are of interest and they underline for which QKD scheme and for which configuration gravity is relevant. Key rates for IA and DR are caracterized by a specific transmission length. We have found that this length is not changed when only Alice, the reference party of the post-processing procedure, is accelerating. If Bob's motion is non-inertial it becomes shorter. In RR the length is virtually infinite in general and it becomes smaller but never negative even when gravity is affecting the two parties. Gravity seems to be more relevant in the case of CA. Indeed, the transmission length is in any case shorter in DR and, furthermore, contrary to IA, the communication length has an upper bound even in RR when the non-reference party has a sufficiently high acceleration. To summarize, in the case of IA from Eve, the performance of the protocols are not significantly changed if gravity is affecting only the reference party. In the case of CA from Eve, this happens only for RR.   
The method described here, used to incorporate the effects of gravity on quantum information protocols, is being generalized to the $3+1$ spacetime and to multimode acceleration in \cite{DD, GD}. QKD with CV would be the natural setting to apply these new developments. Nevertheless, the extension of our present investigation is a difficult problem and will be the matter of a future work.   
\\
\\
One more comment is worth. It is known, that the key rate of CA for RR can be optimized adding properly some noise at Bob's side. This is possible if the noise affects more the correlations between Eve and Bob than the correlations between Alice and Bob. It can be implemented including a beam splitter before Bob's measurement, which combine the signal with a thermal state of variance $N$ and additional noise referred to the input $\chi_B$.
 It is possible to optimize the secret key rate by correctly tuning Bob added noise $\chi_B$ \cite{Gar}.
One might think that the effect of the vacuum thermalization due to Bob's acceleration, known as the Unruh effect, would play the role of $\chi_B$. Unfortunately, this does not work because the described effect is much smaller than the unavoidable modes mismatch between accelerating and inertial wave packets.



\acknowledgments
The author thanks the National Science Centre, Sonata BIS Grant No. DEC-2012/07/E/ST2/01402 for the financial support.
Relevant discussions with Andrzej Dragan, Piotr T. Grochowski and Zehua Tian are recognized.


\begin{thebibliography}{99}
\bibitem{loss}
We have here expressed the trasmittivity $T$ as a function of the losses $l$, $T=10^{-l/10}$.
\end{thebibliography}
\end{document}